\documentclass[aps,pre,twocolumn,showpacs,amsmath,amssymb,groupedaddress]{revtex4-1}

\usepackage{graphicx}
\usepackage[]{lineno}
\usepackage{color}

\begin{document}

\title{Correlations and forecast of death tolls in the Syrian conflict}

\author{Kazuki Fujita}
\affiliation{Department of Physics, Kyoto University, Kyoto 606-8502, Japan}
\author{Shigeru Shinomoto}
\affiliation{Department of Physics, Kyoto University, Kyoto 606-8502, Japan}
\author{Luis E C Rocha}\email{luis.rocha@ki.se}
\affiliation{Department of Public Health Sciences, Karolinska Institutet, Stockholm, Sweden}
\affiliation{Department of Mathematics, Universit\'e de Namur, Namur, Belgium}

\date{\today}

\begin{abstract}
The Syrian civil war has been ongoing since 2011 and has already caused thousands of deaths. The analysis of death tolls helps to understand the dynamics of the conflict and to better allocate resources to the affected areas. In this article, we use information on the daily number of deaths to study temporal and spatial correlations in the data, and exploit this information to forecast events of deaths. We find that the number of deaths per day follows a log-normal distribution during the conflict. We have also identified strong correlations between cities and on consecutive days, implying that major deaths in one location are typically followed by major deaths in both the same location and in other areas. We find that war-related deaths are not random events and observing death tolls in some cities helps to better predict these numbers across the system. 
\end{abstract}


\maketitle

\section{Introduction}
\noindent
The outbreak of the current Syrian armed conflict occurred in March 2011 as a consequence of protests demanding democratic reforms and the end of the current government. These protests quickly escalated and within weeks were widespread in key cities all over Syria, eventually giving rise to groups against and in favor of the government~\cite{Lynch2013}. Since then, the Syrian civil war has been marked by a large number of deaths of both civilians and military personnel. Although a matter of debate~\cite{Taylor2016}, estimates claim that 470,000 people have been killed and at least 3 million refuged or migrated to foreign countries by the end of 2015~\cite{Research2015}.

The dynamics of wars is complex and involves interdependent cultural, ethnic, political and economic variables. Modeling efforts have been employed to forecast the outbreak of conflicts~\cite{Lim2007, Brandt2011,Chadefaux2014, Helbing2015, Parens2016} and to understand their dynamics~\cite{Morgenstern2013, Stauffer2013, Johnson2015}. There is also much interest on estimating the number of casualties and death tolls. Such information helps to allocate resources, estimate the magnitude of the conflict, develop war strategies from the military and political points of view~\cite{Economist2005}, and to quantify the burden of the war on health systems (needed for example to deliver humanitarian aid) and on the society~\cite{Seybolt2013,Sapir2015}. Reliable data on death tolls are difficult to obtain and different methods exist to improve data collection during and after the conflict~\cite{Seybolt2013}. Higher resolution temporal data sets (at daily and weekly resolution) have however became increasingly available in recent years, allowing researchers to employ advanced methods of time-series analysis to make predictions on death tolls~\cite{Rusch2011, Mangion2012, White2016} and to study the dynamics of conflicts~\cite{Jaeger2008,Haushofer2010, Kurzman2014}.

In this article, we study the daily time series of death tolls in the current Syrian civil war and look for the possibility of detecting signs of war-related tragic events based on temporal correlation within individual cities and spatial correlation across different cities. We compare these results to the statistics of a benchmark country, England, which is a representative country not undergoing domestic armed conflicts, in which the daily number of deaths at different cities is expected to have no direct causal relation. 

We identify that the number of war-related deaths during the conflict follows a log-normal distribution whereas the number of deaths for English cities is better described by the normal distribution. We then first investigate the correlations in the number of deaths between cities and find that in Syria this number is not random but follows temporal patterns that may be used to forecast death tolls. Secondly, we perform simulations by assimilating models to the characteristics of the real data such as slow non-stationary fluctuations or rapid daily correlations within each city, and examine the extent to which the observed temporal and inter-city correlations are explained by these apparent characteristics. Thirdly, we carry out the Granger causality analysis~\cite{Granger1969, Honerkamp1998} to see if there are statistical causal relations across different cities. Finally, we attempt to predict the number of deaths; given a set of data for each country, we fix the parameters of prediction models using the first half of the time series and then apply the models to the latter half to see if the models give a certain prediction on the future. The predictability depends not on the model but essentially on whether there is statistical causal relation in each data set. We find for Syrian data that the vector auto-regression (VAR)~\cite{Hamilton1994} model that takes account of inter-city correlations outperforms the auto-regression (AR)~\cite{Hamilton1994} model that uses only single time series of individual cities separately. This indicates that the information of war-related events occurring in some cities may be used for warning of potential occurrence of future events in other cities.

\section{Death tolls}

\begin{figure}[thb]
\centering
\includegraphics[scale=1.0]{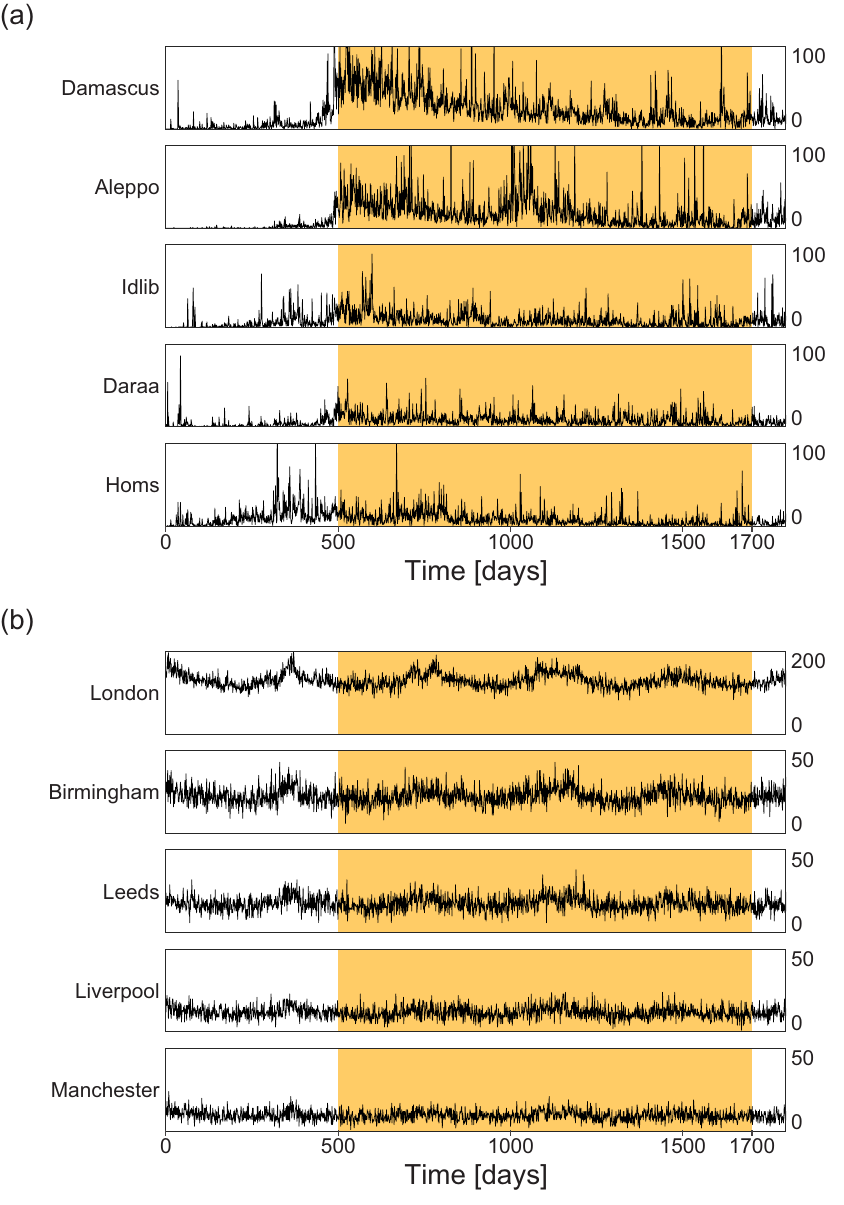}
\caption{\textbf{Daily time series of the number of deaths.} The figure shows the number of deaths per day in different cities of (a) Syria and (b) England. The highlighted interval shows the time period of 1200 days in which we have performed the correlation and the Granger causality analyses. For the prediction analysis, the first half of this period is used to fix the prediction models, and the latter half is used to examine the predictability.}
\label{fig01}
\end{figure}

The data used in this study come from The Violations Documentation Center (VDC) in Syria (www.vdc-sy.info). The VDC has been collecting information on death tolls in the Syrian civil war since June 2011 and retrospectively since the outbreak of the conflict in March 18, 2011 (our time zero). We aggregate data at the province level, adding together adults, children, civilians and military personnel to get a unified number of deaths per day. Figure~\ref{fig01}(a) shows the time-series of death tolls in the top 5 provinces with most casualties, i.e.\ Damascus (including the suburbs), Aleppo, Idlib, Daraa and Homs. We focus our analysis in the times after the shock, starting at about 500 days from the outbreak of the conflict, and during 1200 days. 

The office for National Statistics (www.ons.gov.uk) provides the daily number of deaths in England, that we use as a reference, from 2010 to 2014. We select 5 large cities in England: the greater London area (hereafter referred simply as London), Birmingham, Leeds, Liverpool and Manchester. We also aggregate adults and children in this case. Figure~\ref{fig01}(b) shows a strong seasonal pattern in which deaths are more common during winter in all studied cities. 

\begin{figure}[thb]
\centering
\includegraphics[scale=1.0]{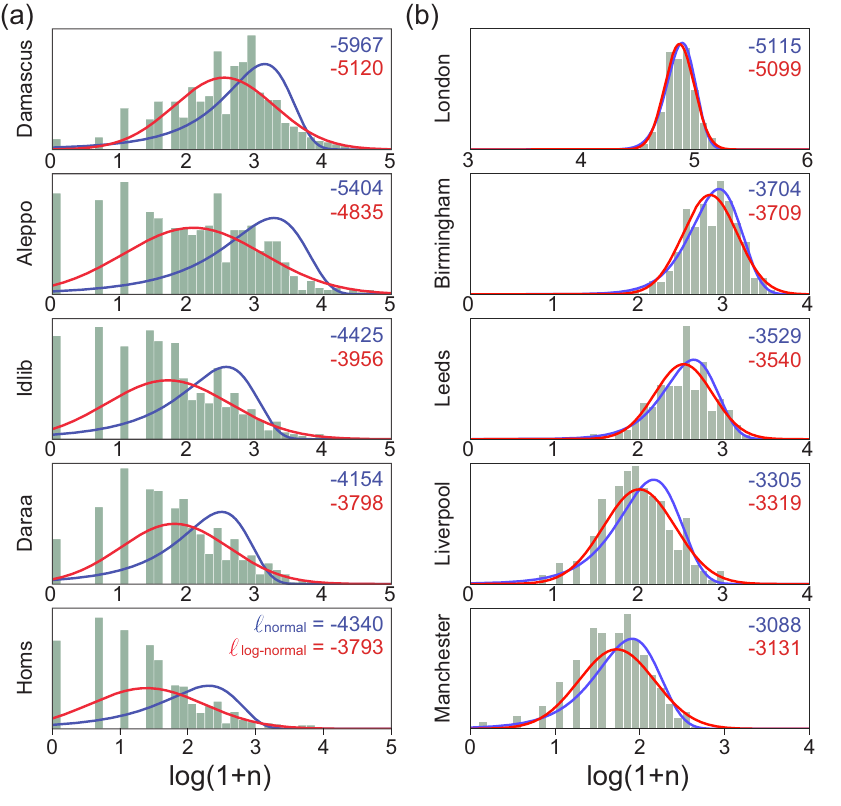}
\caption{\textbf{Distribution of number of deaths per day.} (a) Syrian and (b) English cities. Distributions of $x = \log (1+n)$ are plotted. The log-normal and normal distributions fitted to each data are displayed on top of the histograms, with their goodness-of-fit represented in terms of the log-likelihood $\ell = \log L$.}
\label{fig02}
\end{figure}

The distribution of the number of deaths per day is modeled with a parametric distribution function. The log-normal distribution means that it is relatively often to observe large number of deaths per day in comparison to the typical values. On the other hand, the normal distribution means that the number of deaths per day fluctuates evenly around the mean value. The model goodness-of-fit was corroborated by comparing their log-likelihood $\ell \equiv \log L$. We find that the distribution follows a log-normal distribution for all cities in Syria and London in England (Fig.~\ref{fig02}). Though other cities in England exhibited slightly larger likelihood for the normal distribution, Syrian data in question exhibited absolutely large likelihood, and accordingly we analyze all the time series (including English data) using the transformed variables $x = \log (1+n)$. Note that the results are essentially unchanged even if we take the raw numbers $x = n$ for English data.

\section{Correlation Analysis}

\begin{figure}[thb]
\centering
\includegraphics[scale=1.0]{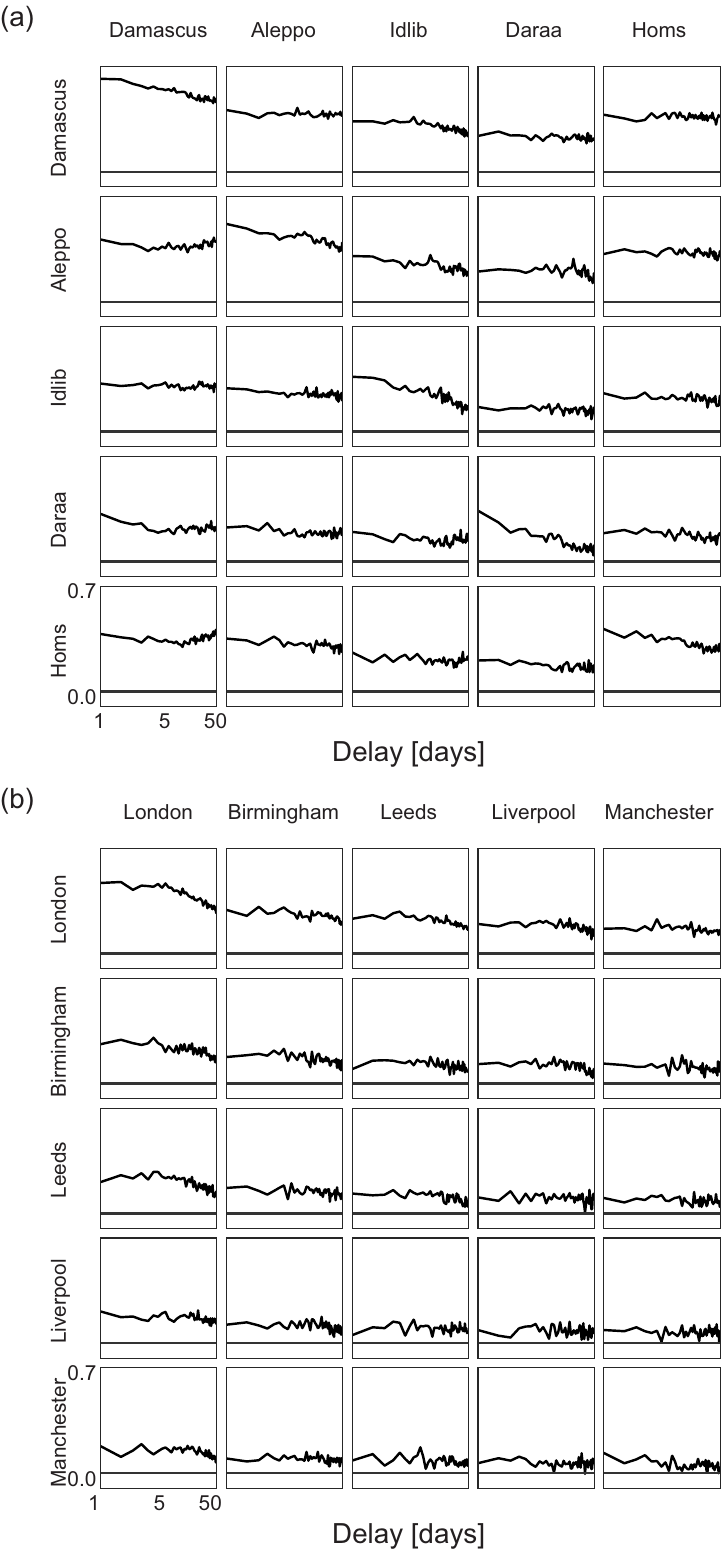}
\caption{\textbf{Correlation analysis.} Temporal and spatial correlation represented by the auto- and cross-correlation between cities given by $\{ \phi_{ij}(t) \}$ in (a) Syria and (b) England. The delay $t$ in the abscissa is represented in logarithmic scale. }
\label{fig03}
\end{figure}

In this section, we analyze the temporal structure of the daily deaths in individual cities and the spatial correlation across cities in the same country. 
The cross-correlation of the daily data between $i$th and $j$th cities is given as
\begin{equation}
\label{eq01}
\phi_{ij}(t) = 
\frac
{\displaystyle\frac{1}{T-t}
\displaystyle\sum_{s=1}^{T-t}(x_{i}(s + t)-\bar{x_{i}})
(x_{j}(s)-\bar{x_{j}})
}
{\displaystyle\frac{1}{T}
{\Biggl(
\Bigl(
\displaystyle\sum_{s=1}^{T}
{(x_{i}(s)-\bar{x_{i}})}^{2}
\Bigr)
\Bigl(
\displaystyle\sum_{s=1}^{T}
{(x_{j}(s)-\bar{x_{j}})}^{2}
\Bigr)
\Biggr)}^{\frac{1}{2}}}
\end{equation}
where $t$ is the time difference measured in the unit of day, $T=1200$, and $\bar{x}_i \equiv \sum_{t=1}^T x_i(t)/T$. The auto-correlation in the $i$th city is given by $\phi_{ii}(t)$. The correlation functions $\{ \phi_{ij}(t) \}$ computed for the entire time series of Syria and England are displayed in Figs.~\ref{fig03} (a) and (b), respectively.

\subsection{Temporal correlation}

The diagonal elements in Fig.~\ref{fig03} represent auto-correlations in individual cities, $\{\phi_{ii}(t)\}$. For the Syrian data, the prominent positive auto-correlation lasting for a few days, particularly in Damascus, Aleppo and Homs, suggests that high death tolls in one day is followed by high death tolls on the next day in the same city, possibly reflecting the situation that individual war-related events have caused a number of deaths in the subsequent days or that major attacks triggers a series of new attacks. Note that auto-correlation is also present in English cities (in London in particular), in which internal armed conflicts are absent. The auto-correlation of long timescale in this case may reflect the seasonal modulation, which is seen in Fig.~\ref{fig01}(b).

\subsection{Spatial correlation}

The spatial, inter-city or cross-correlation is represented as the off-diagonal elements in Fig.~\ref{fig03}. We may observe significant positive cross-correlation across Syrian cities, in particular from Damascus to Aleppo $\phi_{21}(t)$, $(t\ge1)$, from Damascus to Idlib $\phi_{31}(t)$, and from Damascus to Homs $\phi_{51}(t)$, suggesting that high death tolls in Damascus is followed by high death tolls in Aleppo, Idlib, and Homs (Fig.~\ref{fig03}(a)). Similarly, significant cross-correlation is observed from Homs to Aleppo $\phi_{25}(t)$, from Aleppo, Idlib and Homs to Damascus $\phi_{12}(t)$, $\phi_{13}(t)$ and $\phi_{15}(t)$, and from Aleppo to Homs $\phi_{52}(t)$. Note that the cross-correlation are present also in some cases in England, though at much lower intensity and not exhibiting rapid changes as in Syria (Fig.~\ref{fig03}(b)).

\section{Simulated Cities}

We have seen in Fig.~\ref{fig01} that non-stationary fluctuations of long timescale are taking place commonly across different cities in each country; the death tolls in Syria have been slowly decreasing on average (Fig.~\ref{fig01}(a)), while those in England exhibit seasonal modulation (Fig.~\ref{fig01}(b)). To examine the extent to which the observed temporal and spatial correlations are explained by these slow fluctuations, we create the following assimilated time series for each city and perform the same correlation analysis. 

\begin{table}[t]
\centering
\begin{tabular}{|l|c|c|}
\hline
Syrian Cities \, & \, Mean \, & \, Variance \, \\
\hline
Damascus & 3.03 & 0.83 \\
Aleppo & 2.60 & 1.01 \\
Idlib & 1.99 & 0.89 \\
Daraa & 2.03 & 0.75 \\
Homs & 1.84 & 0.91 \\
\hline\hline
English Cities \, & \, Mean \, & \, Variance \, \\
\hline
London & 4.86 & 0.13 \\
Birmingham & 3.12 & 0.23 \\
Leeds & 2.89 & 0.26 \\
Liverpool & 2.48 & 0.32 \\
Manchester & 2.3 & 0.33 \\
\hline
\end{tabular}
\caption{\textbf{Mean and variance of the number of deaths in Syrian and English cities.} The number of deaths is measured as $x=\log(1+n)$.}
\label{tab01}
\end{table}

\begin{figure}[thb]
\centering
\includegraphics[scale=1.0]{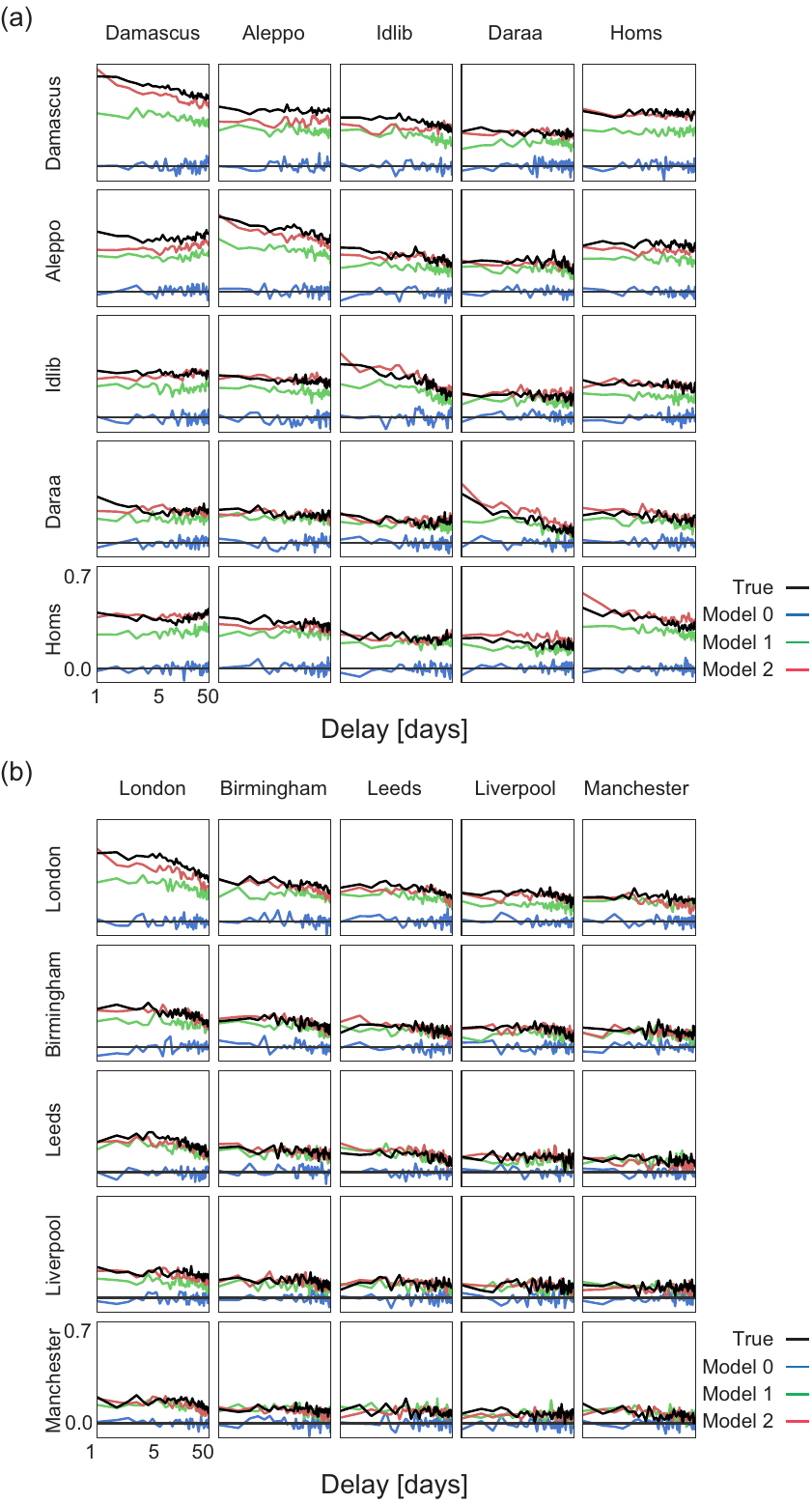}
\caption{\textbf{Correlation analysis for simulated data} (a) and (b) Correlations for time series constructed by assimilating Syrian and English cities, respectively. Model 0 (blue curves) represents a stationary time series of independent Gaussian random numbers, given the mean and variance in each city. Model 1 (green curves) takes account of slow non-stationary modulation in each country. Model 2 (red curves) takes account of daily correlation. The memory parameters were chosen as $h_{\text{Damascus}}=0.3, h_{\text{Aleppo}}=0.2, h_{\text{Idlib}}=0.2, h_{\text{Daraa}}=0.2, h_{\text{Homs}}=0.2$, $h_{\text{London}}=0.2, h_{\text{Birmingham}}=0, h_{\text{Leeds}}=0, h_{\text{Liverpool}}=0, h_{\text{Manchester}}=0$.}
\label{fig04}
\end{figure}

\noindent
\textbf{Model 0: Stationary uncorrelated time series.}
We first generate a series of independent Gaussian random numbers $\xi_i(t)$, given the mean and variance of each ($i$th) city (see Table~\ref{tab01}):
\begin{equation}
x^{0}_i(t) = \xi_i(t).
\end{equation}

\noindent
\textbf{Model 1: Non-stationary uncorrelated time series.}
We modulate the stationary time series Model 0 according to the slow modulation observed in each city, which can be obtained by smoothing the original data with the Gaussian kernel with the standard deviation of $k=10$ days,
\begin{equation}
\Delta \tilde{x}_i(t) = \sum_{s=1}^{T} x_i(s) \frac{1}{\sqrt{2 \pi} k} e^{-\frac{(t-s)^2}{2k^2}} - \bar{x}_i. \label{slowfluctuation}
\end{equation}
The smoothed modulation is added to the time series of Model 0 as
\begin{equation}
x^{1}_i(t) = x^{0}_i(t) + \Delta \tilde{x}_i(t).
\end{equation}
This addition in terms of the logarithmic coefficient $x=\log (1+n)$ corresponds to multiplying the modulation to the original number of deaths.

\noindent
\textbf{Model 2: Non-stationary correlated time series.} 
The strong temporal correlation $\phi_{ii}(t)$ lasting for a few days in Syrian cities may be reproduced by adding memory $h_i$ to the random variable $y_i(t)$, such that 
\begin{equation}
y_i(t)=(1-h_i) \xi_i(t)+ h_i y_i(t-1).
\end{equation}
We may add the higher order memory terms if needed. A stationary correlated time series may be constructed by iterating this equation. By adding the slow fluctuation $\Delta \tilde{x}_i(t)$ to the stationary time series $y_i(t)$, we can obtain a non-stationary correlated time series:
\begin{equation}
x^{2}_i(t) = y_i(t) + \Delta \tilde{x}_i(t).
\end{equation}

Figure~\ref{fig04} summarizes the results of the correlation analysis applied to Models 0, 1, and 2 in reference to the real data. As expected, the uncorrelated stationary time series Model 0 rarely exhibited significant correlation. Model 1 that adopted the slow modulation has reproduced the most part of slow correlations in English data, but has not succeeded in reproducing the strong auto-correlation and cross-correlation in Syrian data. Model 2 was able to reproduce the strong auto-correlation of the Syrian data by suitably accommodating the memory parameter $h_i$ (Fig.~\ref{fig04}). Nevertheless, the strong correlations across real Syrian cities were not reproduced even with this Model 2. The result implies that some inter-city correlations are not spurious, but death tolls are really correlated between some cities such that, for example, an increase of death tolls in Damascus is followed by an increase of death tolls in Aleppo, Daraa and Homs, or an increase in Aleppo is followed by an increase in Homs.

\section{Granger causality}

\begin{figure}[thb]
\centering
\includegraphics[scale=1.0]{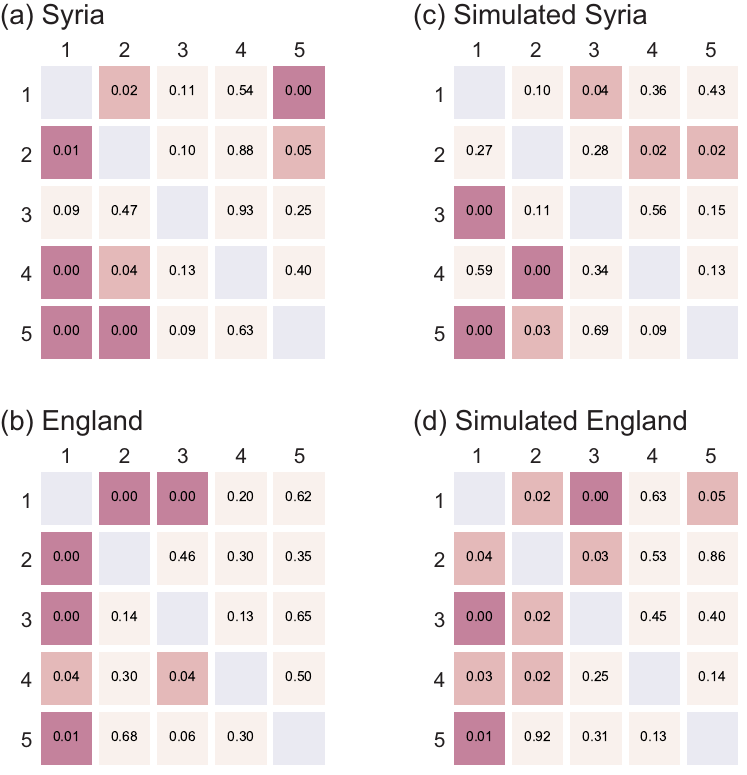}
\caption{\textbf{Granger causality analysis applied to the raw data.} (a) and (b) $p$-values of Granger causality for real Syrian and English data, respectively. (c) and (d) $p$-values of Granger causality for the simulated data given by Model 2. For Syrian data: 1: Damascus; 2: Aleppo; 3: Idlib; 4: Daraa; 5: Homs. For English data: 1: London; 2: Birmingham; 3: Leeds; 4: Liverpool; 5: Manchester.}
\label{fig05}
\end{figure}

Here we try to detect statistical causal relation between different cities in both countries, using a standard pairwise-conditional Granger causality (GC) analysis~\cite{Granger1969, Honerkamp1998}. The naive GC analysis indicated many inter-city correlations not only in Syria (Figs.~\ref{fig05}(a)) but also in England (Figs.~\ref{fig05}(b)), although the direct causal relations between English cities are expected to be absent. The GC analysis is known to be vulnerable to non-stationary fluctuations and likely to suggest spurious correlations~\cite{Granger1988}. Actually, by applying the same GC analysis to the assimilated data (Model 2), we also obtained similar correlations (Figs.~\ref{fig05}(c) and (d)). The result suggests that the seasonal modulation has induced the virtual causality.

\begin{figure}[thb]
\centering
\includegraphics[scale=1.0]{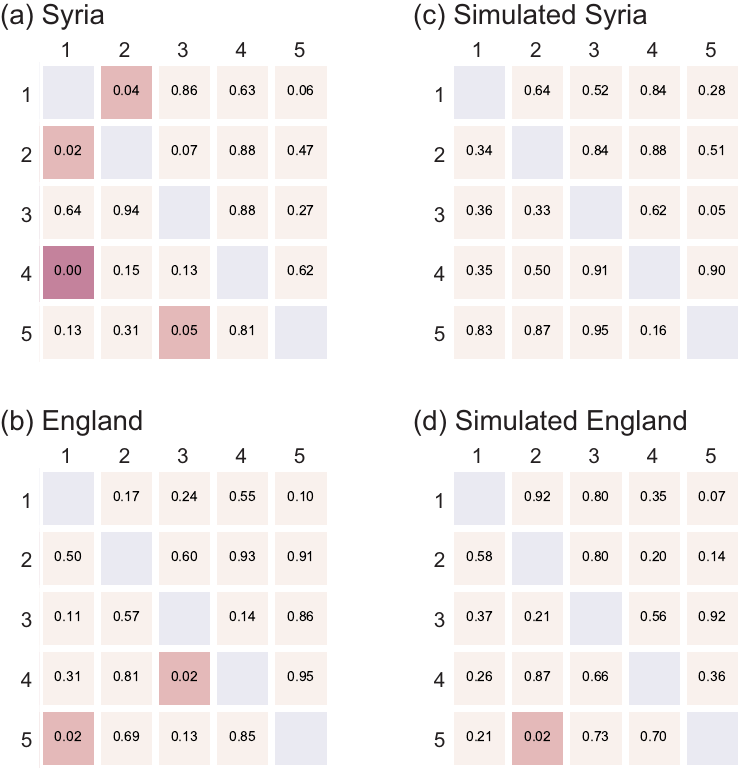}
\caption{\textbf{Granger causality analysis applied to the time difference data $z_i(t) \equiv x_i(t)-x_i(t-1)$.} (a) and (b) $p$-values of Granger causality for real Syrian and English data, respectively. (c) and (d) $p$-values of Granger causality for simulated data given by Model 2. For Syrian data: 1: Damascus; 2: Aleppo; 3: Idlib; 4: Daraa; 5: Homs. For English data: 1: London; 2: Birmingham; 3: Leeds; 4: Liverpool; 5: Manchester. }
\label{fig06}
\end{figure}

It is known that the influence of slow non-stationary fluctuations to the GC analysis may be mitigated by analyzing the temporal difference~\cite{Hamilton1994}, 
\begin{equation}
z_i(t) \equiv x_i(t)-x_i(t-1).
\end{equation}
The operation succeeded in removing spurious inter-city correlations from not only the simulated Model 2 (Fig.~\ref{fig06} (c) and (d)), but also from real English data (Fig.~\ref{fig06}(b)). However, the real Syrian data is left with directed inter-city correlations from Damascus to Aleppo and Daraa, from Aleppo to Damascus, and from Idlib to Homs (Fig.~\ref{fig06}(a)). Thus this analysis also suggests that death tolls are dependent across those Syrian cities.

\section{Forecast}

In this section, we attempt to predict the death tolls in Syria and England, using the auto-regression (AR) model (Eq.~\ref{eq02}) and the vector auto-regression (VAR) model (Eq.~\ref{eq03})~\cite{Hamilton1994}. We use the first half of the time series ($600$ days) to fix the model parameters, and apply the models to the latter half to see if the models give efficient prediction on the future number of deaths in each country. 

The AR model only uses information of a single time series to forecast values of the corresponding time series, as given by 
\begin{eqnarray}
\hat{x}_i(t) = c + \sum_{s=1}^{m} a_i(s) x_i(t-s) + \epsilon(t), \label{eq02}
\end{eqnarray}
where $\hat{x}_i(t) $ represents the predicted value, $\{a_i(s)\}_{s=1, \cdots, m}$ are parameters determined with the temporal correlation $\phi_{ii}(s)$ (Eq.~\ref{eq01}) computed for the first half of the time series, and $m$ is the order of regression. We take $m=5$ for the current analysis.

The VAR model uses the time series of all cities simultaneously to forecast values of all cities at once, as given by
\begin{eqnarray}
\hat{\mathbf{x}}(t) = \mathbf{c} + \sum_{s=1}^{m} \mathbf{A}(s) \mathbf{x}(t-s) + \mathbf{\epsilon}(t), \label{eq03}
\end{eqnarray}
where $\mathbf{x}(t)$ represents a vector comprising of five cities $(x_1(t), x_2(t), \cdots, x_5(t))^{t}$, and $\mathbf{A}(s)$ is a matrix whose elements are determined with the temporal correlation $\{\phi_{ij}(s)\}$ (Eq.~\ref{eq01}).

If there are correlations between cities, there is room for the VAR model to make the better forecast in comparison to the AR model that only uses information of a single time series. As a reference, these two models are compared with simpler models: (i) predicting the value simply with that of the preceding date, and (ii) predicting with a single fixed value given by averaging over the past first half time series.

\begin{figure}[thb]
\centering
\includegraphics[scale=1.0]{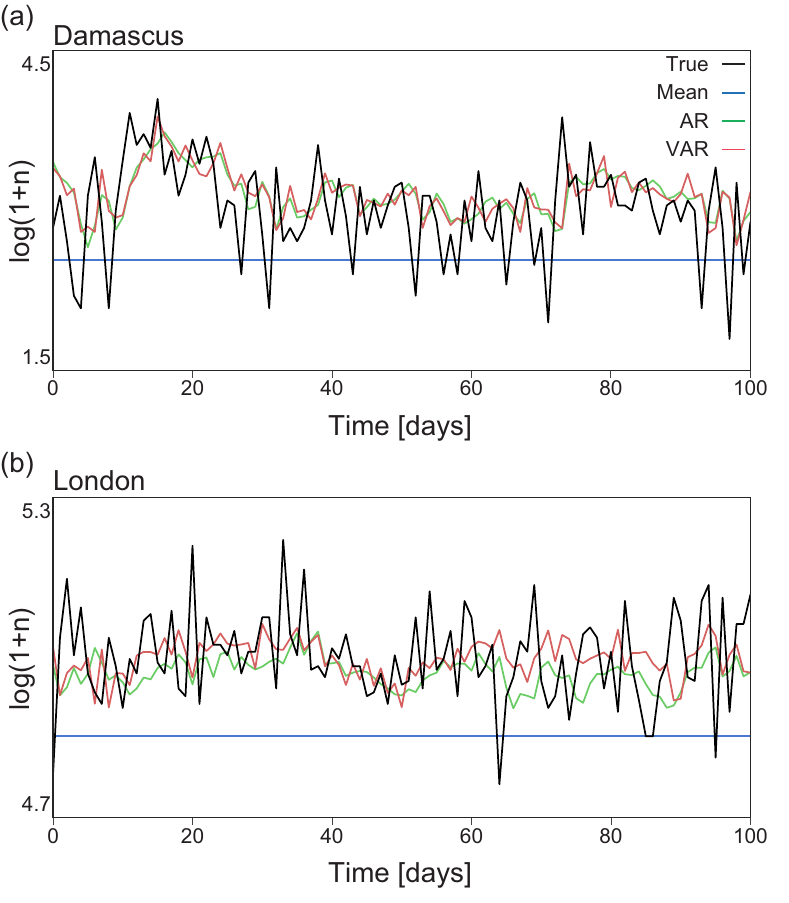}
\caption{\textbf{Prediction of the number of deaths.} The plots show the initial 100 days of the latter half of the time series, which is used for evaluating the predictability of the models. (a) Syrian city of Damascus and (b) English city of London. The ordinates are in logarithmic scale. Predictions made by the models (ii) Average (blue curve), (iii) AR (green curve), and (iv) VAR (red curve) are plotted on top of the real time series (black curve). .}
\label{fig07}
\end{figure}

\begin{table*}[t]
\centering
\begin{tabular}{|l|c|c|c|c|c|}
\hline
& \, Preceding \, & \, Average \, & \,\,\, AR \,\,\, & \,\, VAR \,\, & \, AR-VAR \, \\
\hline\hline
Damascus & 0.717 & 0.552 & 0.500 & 0.478 & $+0.022$ \\
Aleppo & 1.383 & 1.064 & 1.016 & 1.006 & $+0.010$ \\ 
Idlib & 1.295 & 0.823 & 0.787 & 0.791 & $-0.004$ \\
Daraa & 0.901 & 0.638 & 0.635 & 0.570 & $+0.065$ \\
Homs & 1.173 & 0.711 & 0.831 & 0.732 & $+0.099$ \\
\hline\hline
London & 0.017 & 0.017 & 0.011 & 0.011 & $-0.000$ \\
Birmingham & 0.088 & 0.059 & 0.052 & 0.051 & $+0.001$ \\ 
Leeds & 0.110 & 0.066 & 0.061 & 0.060 & $+0.001$ \\
Liverpool & 0.187 & 0.101 & 0.101 & 0.100 & $-0.001$ \\
Manchester & 0.194 & 0.113 & 0.108 & 0.108 & $-0.000$ \\
\hline
\end{tabular}
\caption{\textbf{Forecast error for Syrian and English cities.} The table shows the mean square error between the forecast provided by each model $\hat{x}_i (t)$ and the true value $x_i (t)$, averaged over the latter half of the time series ($T=600$), $\sum_{t=T+1}^{2T} (\hat{x}_i (t) - x_i (t))^2/T$. ``AR-VAR'' stands for the difference in the prediction errors of the AR and VAR models.}
\label{tab02}
\end{table*}

Figures~\ref{fig07} (a) and (b) demonstrate the predictions made for the Syrian city of Damascus and the English city of London, respectively, using the models (ii) Average, (iii) AR, and (iv) VAR in reference to the real values. Table~\ref{tab02} demonstrates the performances of these four models for the Syrian and English cities in terms of the average prediction error. The AR and VAR models generally perform much better than the other two methods. Among the two regression models, VAR outperforms AR model for the Syrian data; the difference in the prediction errors as represented by ``AR-VAR'' in Table~\ref{tab02} is very large in particular for four cities in Syria, in contrast to the negligible difference in English cities. This result also indicates the presence of significant inter-city correlation in those Syrian cities. 

The total prediction error in the unit of the numbers of dead people is given by 
\begin{equation}
E=\sum_{i=1}^{5} \sum_{t=1}^{T} | n_i(t+T) +1- e^{\hat{x}_i(t+T)} |,
\end{equation}
where $T=600$, $n_i(t)$ is the number of deaths of the $i$th city at day $t$, and $\hat{x}_i(t)$ is the predicted value for $\log (1+n_i(t))$. The difference in the total prediction error between AR and VAR models corresponds to 451 people in the period of 600 days.

\section{Conclusions}

Wars typically cause significant life losses on all sides involved in the conflicts. Estimates of death tolls are important in order to quantify the magnitude of the war, encourage peacemaking and to allocate humanitarian aid. The availability of high-resolution spatiotemporal data on the number of deaths allows researchers to analyze correlations between different cities at different times and to identify trends that could possibly be used to reduce future causalities.

In this paper, we use daily information on the number of deaths in a given city to study spatial and temporal correlations of death tolls in the current Syrian civil war and compared the results with the daily number of deaths in English cities that are not undergoing any domestic conflict. We have explored different models to remove potential virtual correlations in the empirical data, as was the case in English cities mainly due to seasonality. Our analysis showed that significant positive auto-correlation exist in Syrian cities, meaning that days with major number of deaths in a particular city were followed by days with many deaths in the same city, possibly reflecting a sequence of attacks within short periods. Similarly, we have also observed significant cross-correlation (i.e. spatial correlation) between some cities in Syria. This means that deaths in one city were followed by deaths at another city, for example, from Damascus to Aleppo and vice-versa, from Damascus to Daraa, and from Idlib to Homs, possibly reflecting coordinated attacks. Such results are useful since one can exploit them to develop a warning system monitoring the sequence of events taking place at different days and cities aiming to better understand attack strategies.

We have also explored the possibilities to forecast death tolls at different cities. Our analysis have shown that due to the correlations, improved forecast is obtained for Syria if using information from all cities simultaneously in a vector auto-regression model in comparison to single cities in independent auto-regression models. We observe no difference (typically less than $2\%$) for England. Furthermore, for both countries, regression models performed much better than naive methods such as preceding date and the average over the learning period. The important conclusion here is that death tolls can be predicted to a good accuracy and thus such methods could be used to organize the allocation of resources during the conflict.

We finally observe that daily death tolls in Syria follow a log-normal distribution which is in contrast to English cities that generally better follow the normal distribution, though the difference is negligible in the later case. This means that death events are not uniform and events with a large number of deaths are expected during the conflict. We should keep in mind that the numbers of casualties in Syria have been very large and the counting of deaths might have been a very difficult task. Accordingly the counts might have been accompanied with significant errors that could have affected the correlations between specific cities and the overall forecast exercise. Future work should focus on the analysis of different datasets to determine the intensity of the correlations and thus the possibilities for forecast on other contexts.

\acknowledgments
The authors thank Ryota Kobayashi for fruitful discussions. SS and LECR were supported by the Bilateral Joint Research Project between JSPS, Japan, and FRS-FNRS, Belgium. SS was supported by a Grant-in-Aid for Scientific Research from MEXT Japan (26280007) and by JST and CREST. 

\bibliographystyle{unsrt}
\bibliography{main_v5}

\end{document}